\documentclass[twocolumn,superscriptaddress,preprintnumbers,aps,amsmath,amssymb]{revtex4-1}

\usepackage{graphicx} 
\usepackage{color}
\usepackage{tikz}
\usetikzlibrary{arrows}

\begin{document}


\title{Quantum walks assisted by particle number fluctuations}


\author{R. A. Vargas--Hern\'{a}ndez and R. V. Krems}
\affiliation{Department of Chemistry, University of British Columbia, Vancouver, BC V6T 1Z1, Canada}


\date{\today}

\begin{abstract}
{
We study the spreading of a quantum particle placed in a single site of a lattice or binary tree with the Hamiltonian permitting particle number changes. We show that the particle number-changing interactions accelerate the spreading beyond the ballistic expansion limit by inducing off-resonant Rabi oscillations between states of different numbers of particles. We consider the effect of perturbative number-changing couplings on Anderson localization in one-dimensional disordered lattices and show that they lead to decrease of localization. The effect of these couplings is shown to be larger at larger disorder strength, which is a consequence of the disorder-induced broadening of the particle dispersion bands.  
}
\end{abstract}


\maketitle



\section{Introduction}

An important class of quantum computing algorithms is based on quantum walks \cite{QW-QG-QC, QW-review}, the quantum analogue of random walks \cite{QWandRW}. 
Random walks on lattices and graphs are powerful mathematical objects that can be used as algorithmic tools for a variety of problems, including optimization, search and classification.
The efficiency of many such algorithms is determined by `hitting' and `mixing' times, quantifying how long it takes random walks to explore the underlying graphs \cite{QW-graphs}. Depending on the Hamiltonian, quantum walks can be accelerated by dynamical interferences and have potential to offer polynomial or, for some problems, exponential, computation speedup \cite{QW-decision-trees}. 
The role of interferences in quantum walks is perhaps best exemplified by the ballistic expansion of a quantum particle with time $T$ in a periodic lattice \cite{QW-QG-QC,QW-speedup,QW-search,QW-search2}, leading to the $\propto T$ growth of the probability distribution, compared to the $\propto \sqrt{T}$ expansion of the classical random walk. 
Quantum walks have been proven to offer the $\sqrt N$ speed-up of spatial search over $N$ items arranged in a $d$-dimensional lattice, with $d > 4$ \cite{ChildsPRA70}.


With recent advances in the experiments on controlling atoms \cite{atoms-in-optical-lattices_0,atoms-in-optical-lattices_1,atoms-in-optical-lattices_2,atoms-in-optical-lattices_3,atoms-in-optical-lattices_4,Vargas-Krems}, molecules \cite{molecules-in-optical-lattices, molecules-trapped-inside-fullerene-cages_1,molecules-trapped-inside-fullerene-cages_2}, photons \cite{lattice-models-with-photons} and arrays of superconducting qubits \cite{d-wave_1,d-wave_2,d-wave_3,d-wave_4,QW-SQ}, it has become possible to engineer lattice Hamiltonians. This, combined with the importance of the speed of quantum walks for the quantum computing algorithms and for the study of the fundamental limits of the velocity of quantum correlation propagations \cite{group-velocity-QS, gorshkov-monrau-paper}, raises the question if and how lattice or graph Hamiltonians can be engineered to accelerate quantum walks.  The effect of Hamiltonian engineering on quantum walks has been studied in many different contexts. For example, Giraud et al \cite{giraud} showed that Anderson localization impeding quantum walks in disordered systems can be mitigated by adding hopping terms, which provide shortcuts in circular graphs. 
Quantum walks can also be accelerated by coupling a Hamiltonian system to an external bath. 
While the general belief is that particle-environment interactions destroy the coherence of quantum walks leading to transport suppression in ordered systems, multiple recent studies showed that interactions with certain non-Markovian baths provide new pathways for interferences \cite{Prokofev-Stamp,Mohseni-Aspuru,Rebentrost-Aspuru}. The range of particle hopping and particle interactions are also known to determine the speed of quantum information propagation \cite{lr-hopping_1,lr-hopping_2,gorshkov-monrau-paper,Tirtta-Krems}.  

Engineering many-particle (as opposed to single particle) quantum walks is becoming an important research goal \cite{large-scale}. 
As shown by Childs \emph{et al}, quantum walks on a sparse graph can be used to efficiently simulate any quantum circuit \cite{ChildsPRL102} and quantum walks of interacting particles are capable of universal quantum computation \cite{ChildsScience}. Quantum walks of interacting pairs can be used to determine if graphs are isomorphic \cite{Two-QW}.
Particle correlations can be exploited to change the directionality of quantum walks \cite{directionality}. Quantum walks of interacting particles can be used to realize quantum Hash schemes \cite{hash}.  Two-body or multi-particle correlations have been shown to affect quantum walks of few- and many-particle systems in interesting ways \cite{Tirtta-Krems,correlations,correlations2,iso2,correlations3,correlations4,distinguishable,correlations5,interacting-bosons,interacting-particles,discord,binding,directionality, centi}.
These studies consider particle correlations arising either as a consequence of direct density - density interactions or particle quantum statistics. 

Here, we consider an alternative mechanism for accelerating quantum walks, namely quantum walks in a dynamical system governed by a Hamiltonian allowing particle number changes.  Such Hamiltonians can be engineered with quasi-particles, such as excitons \cite{exciton-paper_1,exciton-paper_2,exciton-paper_3,exciton-paper_4}, or with ultracold atoms trapped in optical lattices and immersed in a condensate \cite{bec-paper_1,bec-paper_2}. They are also of significant experimental and theoretical interest due to the relation to the topologically protected states and their possible use in quantum computing \cite{kitaev-model}. In the present work, we show that the particle-number-changing interactions lead to Rabi oscillations, which significantly accelerate the spreading of quantum wave packets in ideal lattices and binary trees. We also consider the effect of such terms on Anderson localization and show that they lead to decrease of the inverse participation ratio in disordered systems. We show that the effect of number-changing interactions on the participation ratio becomes stronger with increasing disorder strength. 

\section{Models}

We consider quantum dynamics governed by the following lattice Hamiltonian:
\begin{eqnarray}
{\cal \hat{H}} &=&  \sum_{i} \omega_i \hat{c}^{\dagger}_{i}\hat{c}_{i} +  t \sum_{\langle i, j \rangle}   \hat{c}^{\dagger}_{j}\hat{c}_{i} \nonumber \\
&&+ v \sum_{\langle i, j \rangle} {c}^{\dagger}_{i} {c}_{i} {c}^{\dagger}_{j} {c}^{}_{j}
+ \hat V_{\rm nc},
\label{model}
\end{eqnarray}
where 
\begin{eqnarray}
\hat V_{\rm nc} =  \Delta \sum_{\langle i, j \rangle} (\hat{c}^{\dagger}_{i}\hat{c}^\dagger_{j} + \hat{c}_{i}\hat{c}_{j}) + \gamma \sum_i (\hat c^\dagger_i + \hat c_i),
\label{pnnc}
\end{eqnarray}
 $\hat c_i$ is the operator that removes the particle from site $i$, the quantities $\omega_i$, $t$, $\Delta$ and $v$ are the Hamiltonian parameters, and the angular brackets indicate that the hopping and interactions are only permitted between nearest neighbour sites. The on-site energy $\omega_i$ is defined as $\omega_i = \Delta \varepsilon + \varepsilon_i$, where $\Delta \varepsilon$ is a constant and $\varepsilon_i$ is varied in the calculations for lattices with on-site disorder (more details below). 
 The term $\hat V_{\rm nc}$ couples different particle-number states. 

The model (\ref{model}) is a special case of the full Hamiltonian for the Frenkel excitons in an ensemble of coupled two-level systems \cite{agranovich}. At $\Delta = 0$, $\gamma = 0$ and $v=0$, this Hamiltonian reduces to the tight-binding model. At $\Delta = 0$ and $\gamma \neq 0$, the model describes the quantum annealer setup of D-wave \cite{d-wave_3}, where currents in interacting superconducting qubits are mapped onto spin states.  

 We consider the few-particle limit of the model (\ref{model}) and calculate the dynamics of quantum walks by diagonalizing the Hamiltonian and constructing the full time evolution operator from the complete set of the corresponding eigenvectors, as was done, for example, in Ref. \cite{Xu-Krems}. In order to describe properly the dynamics governed by the models with $\Delta \neq 0$ and/or $\gamma \neq 0$, the Hilbert space must include multiple particle-number states. We truncate the Hilbert space to include one and three particles for the case $\Delta \neq 0, \gamma = 0$.  When $\Delta = 0, \gamma \neq 0$, the Hilbert space includes the vacuum state (zero particles), one, two, and three particles. 
 As discussed below, this article considers the Hamiltonian parameters, for which the multiple-particle states have high energy. Since the energy of such states increases with the number of particles and the couplings can only change the number of particles by one or two, the contribution of such states decreases with the number of particles. 
  We have verified by a calculations for a lattice with 19 sites that including the states of five particles does not change the results for the Hamiltonian parameters considered here. 

 The on-site energy $\Delta \varepsilon + \varepsilon_i$ determines the energy separation between states with different numbers of particles. 
Throughout this work, we consider the limit $\Delta, \gamma \ll \Delta \varepsilon$. For ideal lattices, $\varepsilon_i = 0$. For disordered lattices, $\varepsilon_i$ is drawn from a uniform distribution of random numbers. In this limit, the state corresponding to one particle at zero time becomes weakly dressed with higher particle-number states. The effect of the dressing can be accounted for by the Schrieffer-Wolf transformation \cite{sw}, which in first order leads to the appearance of 
next-nearest-neighbour hopping terms, as shown in Appendix \ref{ap:SW}. Including higher order terms resulting from the transformation induces longer-range hopping.  The particle can effectively hop by undergoing virtual transitions to higher particle-number states and back. Note that in models with $\gamma \neq 0$, the particle can also hop by virtual transitions to the vacuum state (the state of no particles) and back. 

\subsection{Ideal 1D lattices}

We first consider the well-studied problem of ballistic spreading in an ideal one-dimensional (1D) lattice. At $\Delta = 0$ and $v=0$, a particle placed in an individual lattice site expands as shown by the solid black line in Figure 1. This spreading is much faster than the expansion of the area covered by the classical random walk, illustrated in Figure 1 by the dotted curve. 
Figure 1 shows that the quantum dynamics of a single particle initially placed in a single lattice site is drastically different from both the random walk result and the ballistic spreading
when governed by  the model (\ref{model}) with $\Delta \neq 0$. In particular, the width of the wave packet oscillates at short times, approaching the ballistic-expansion-like behaviour at long times. These calculations are performed for the 1D lattice with $N = 41$ lattice sites with open boundary conditions. As can be seen from Figure 1, the effect of the boundaries is not important until time reaches $\approx 11~t^{-1}$.

In order to understand the origin of the oscillations, we plot in the insets of Figure 1 the average number of particles $\langle n \rangle$ as a function of time. It can be seen that $\langle n \rangle$ oscillates with the same period as the wave packet size. We thus conclude that the oscillations observed in Figure 1 are due to off-resonant Rabi flopping between the state of one particle and the states of multiple particles induced by $\hat V_{\rm nc}$. Figure 1 shows that these coherent oscillations accelerate quantum walks beyond the ballistic limit.  Note that $\langle n \rangle$ in Figure 1 is an average of one and three particles. For $\Delta \varepsilon / t = 20$, $\langle n \rangle < 1.2$, which illustrates that the three-particle subspace remains largely unpopulated at all times.

\begin{figure}[h!]
  \centering
\includegraphics[width=0.5\textwidth]{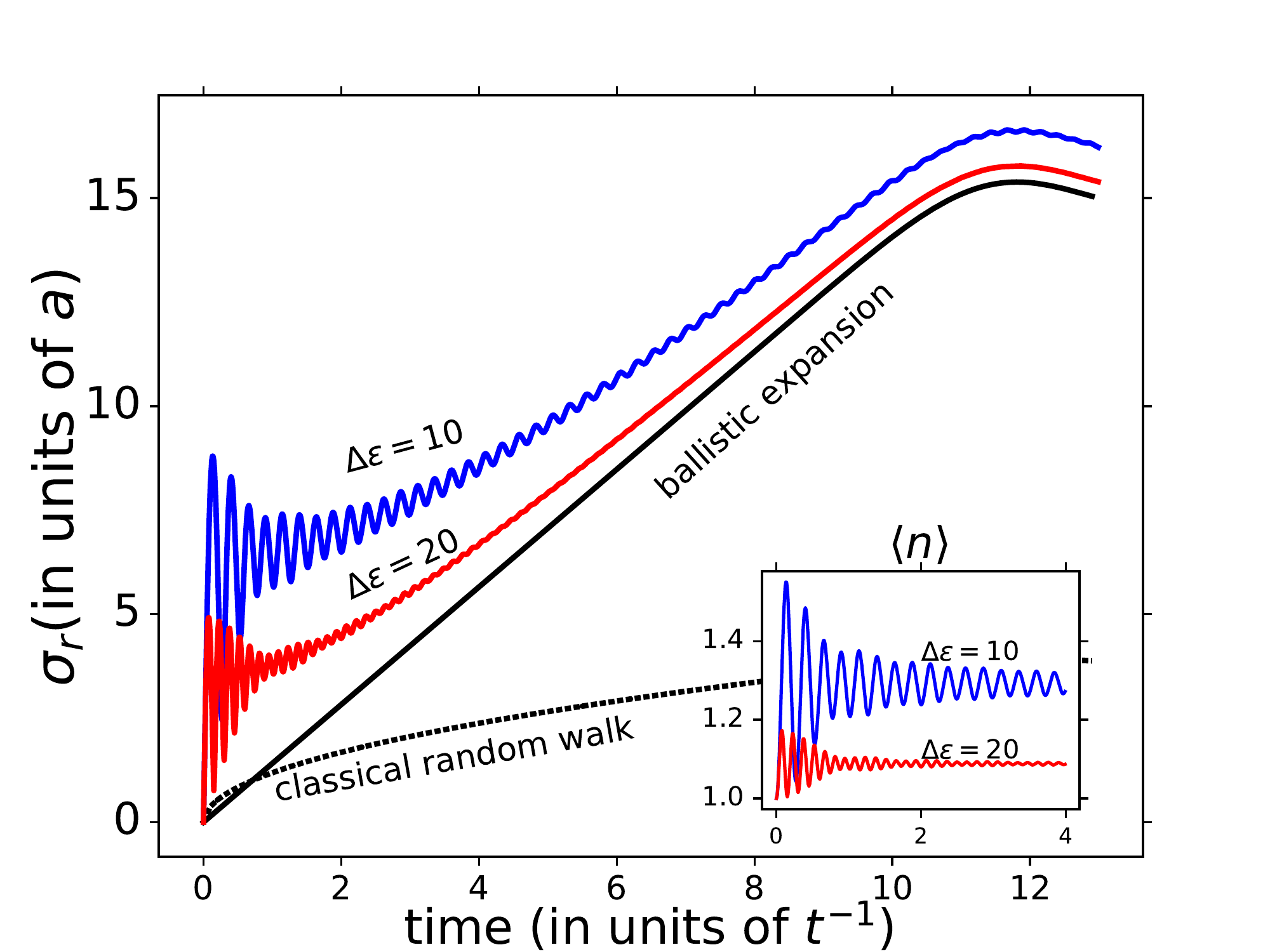} \\
\includegraphics[width=0.5\textwidth]{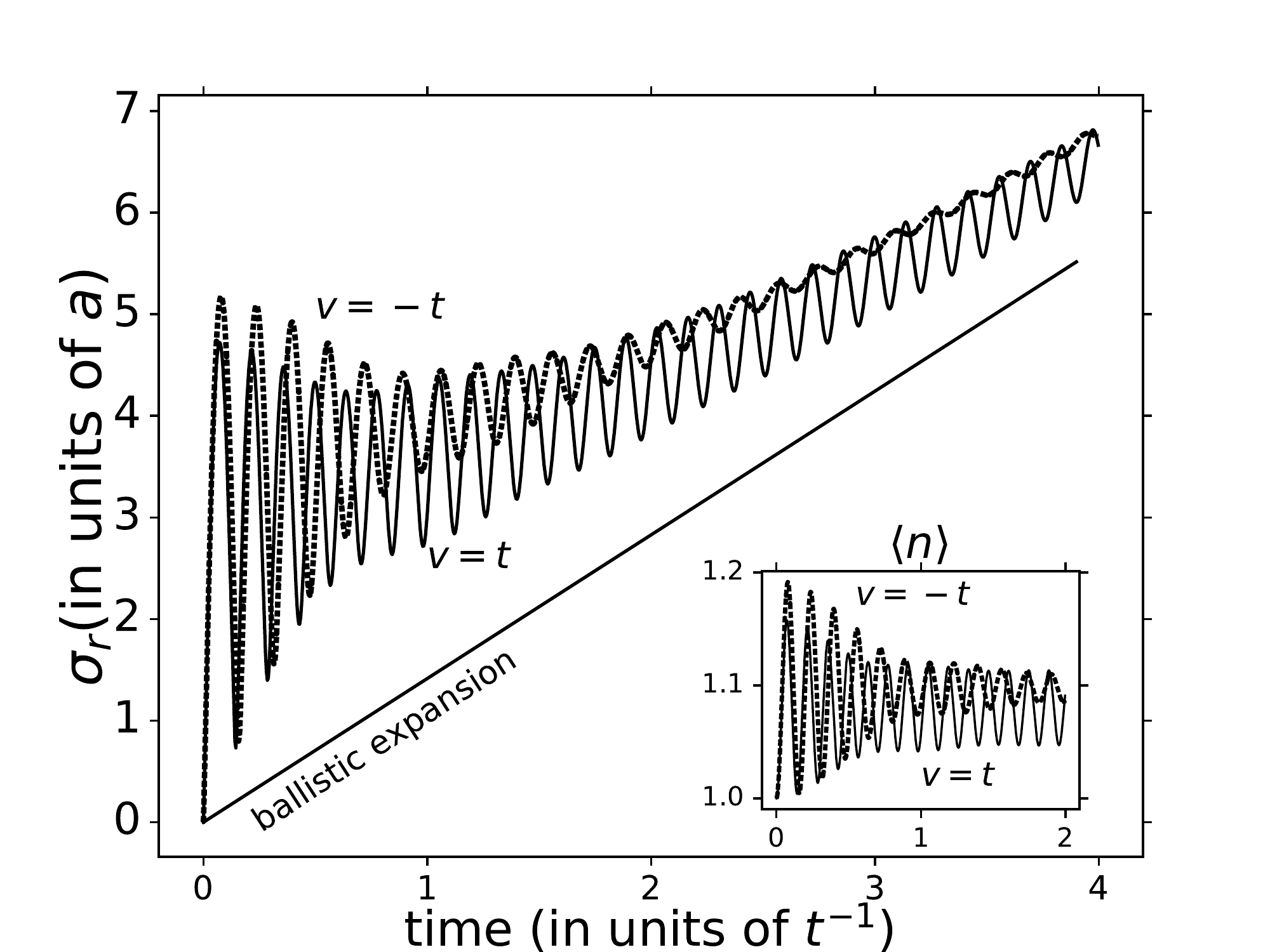}
    \caption{
    Time dependence of the standard deviation (in units of lattice constant) of the wave packet for a particle initially placed in a single site of a one-dimensional ideal lattice. The solid black curves represent the ballistic expansion governed by the Hamiltonian (\ref{model}) with $v=0$ and $\hat V_{\rm nc} = 0$.
    Upper panel: The oscillating curves show the size of the wave packets governed by the Hamiltonian (\ref{model}) with $v=0$, $\gamma = 0$, $\Delta/t = 1$, $t = 1$ and two values of  $\Delta \varepsilon$:   $\Delta \varepsilon = 10/t$ (blue) and $\Delta \varepsilon = 20/t$ (red). 
 Lower panel: The oscillating curves show the size of the wave packets governed by the Hamiltonian (\ref{model}) with $v=\pm1$, $\Delta/t = 1$, $t = 1$ and $\Delta \varepsilon = 20/t$. 
  The insets show the average number of particles $\langle n \rangle$ as a function of time for the corresponding Hamiltonian parameters. Notice that for $\Delta \varepsilon = 20/t$, $\langle n \rangle$ stays below 1.2 at all times. 
}
    \label{fig:sigma_vs_time}
\end{figure}

\begin{figure}[h!]
  \centering
\includegraphics[width=0.5\textwidth]{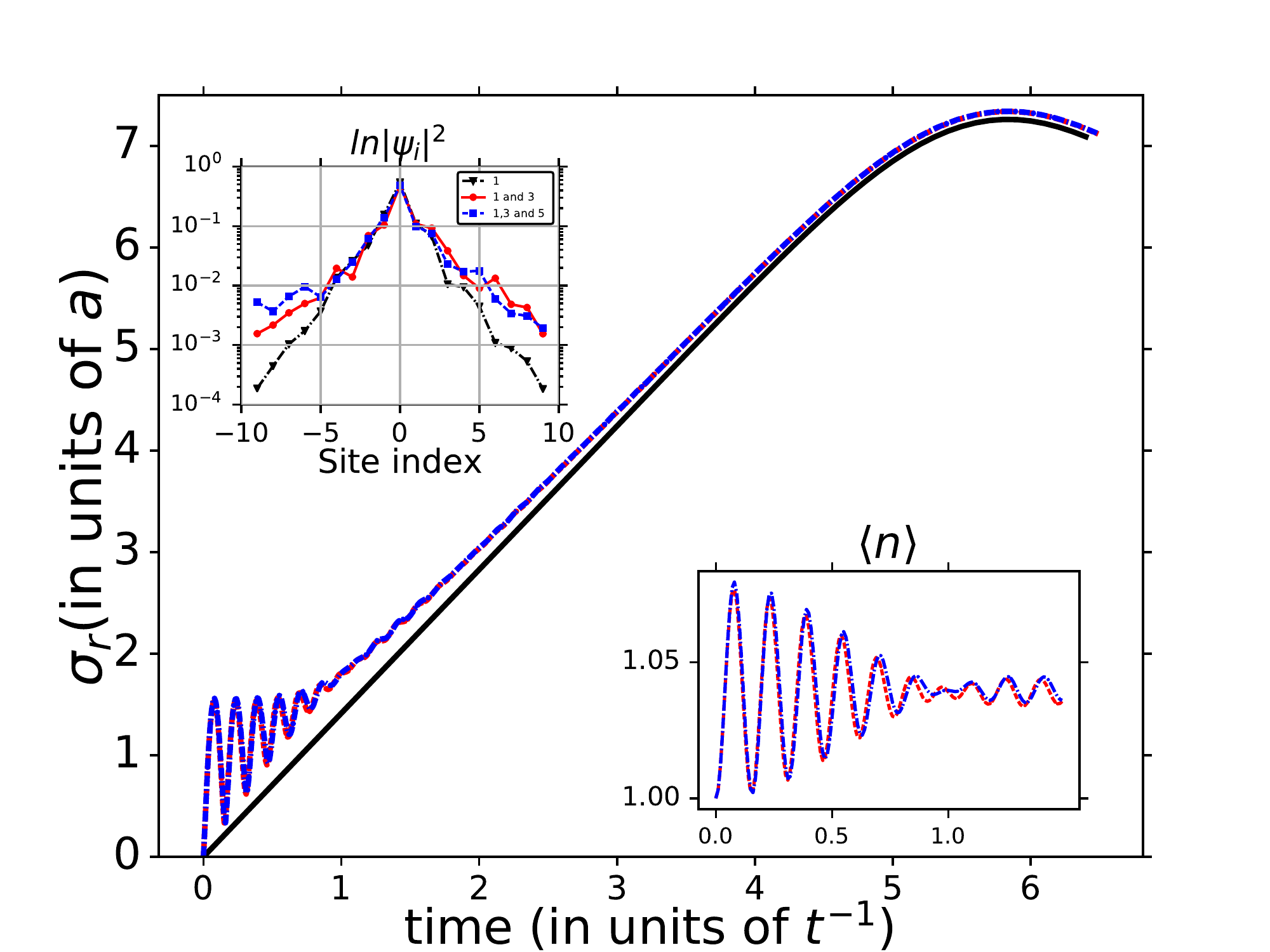} 
\caption{Time dependence of the standard deviation (in units of lattice constant) of the wave packet for a particle initially placed in a single site of a one-dimensional ideal lattice. The solid black curves represent the ballistic expansion governed by the Hamiltonian (\ref{model}) with $v=0$ and $\hat V_{\rm nc} = 0$.
The oscillating curves show the size of the wave packets governed by the Hamiltonian (\ref{model}) with $v=0$, $\gamma = 0$, $\Delta/t = 1$, $t = 1$, $\Delta \varepsilon = 20/t$ and a Hilbert space with different number of particles, 1 particle (black), 1 and 3 particles (red) and 1,3 and 5 (blue). The upper inset shows the logarithm of the particle probability distributions in a disordered 1D lattice with 19 sites. The results are averaged over 50 realizations of disorder and are time-independent with $w = 10/t$. And the lower inset depicts the average number of particles $\langle n \rangle$ as a function of time for the three different wave packets. 
}
    \label{fig:sigma_vs_time_np}
\end{figure}

\begin{figure}[h!]
  \centering
\includegraphics[width=0.5\textwidth]{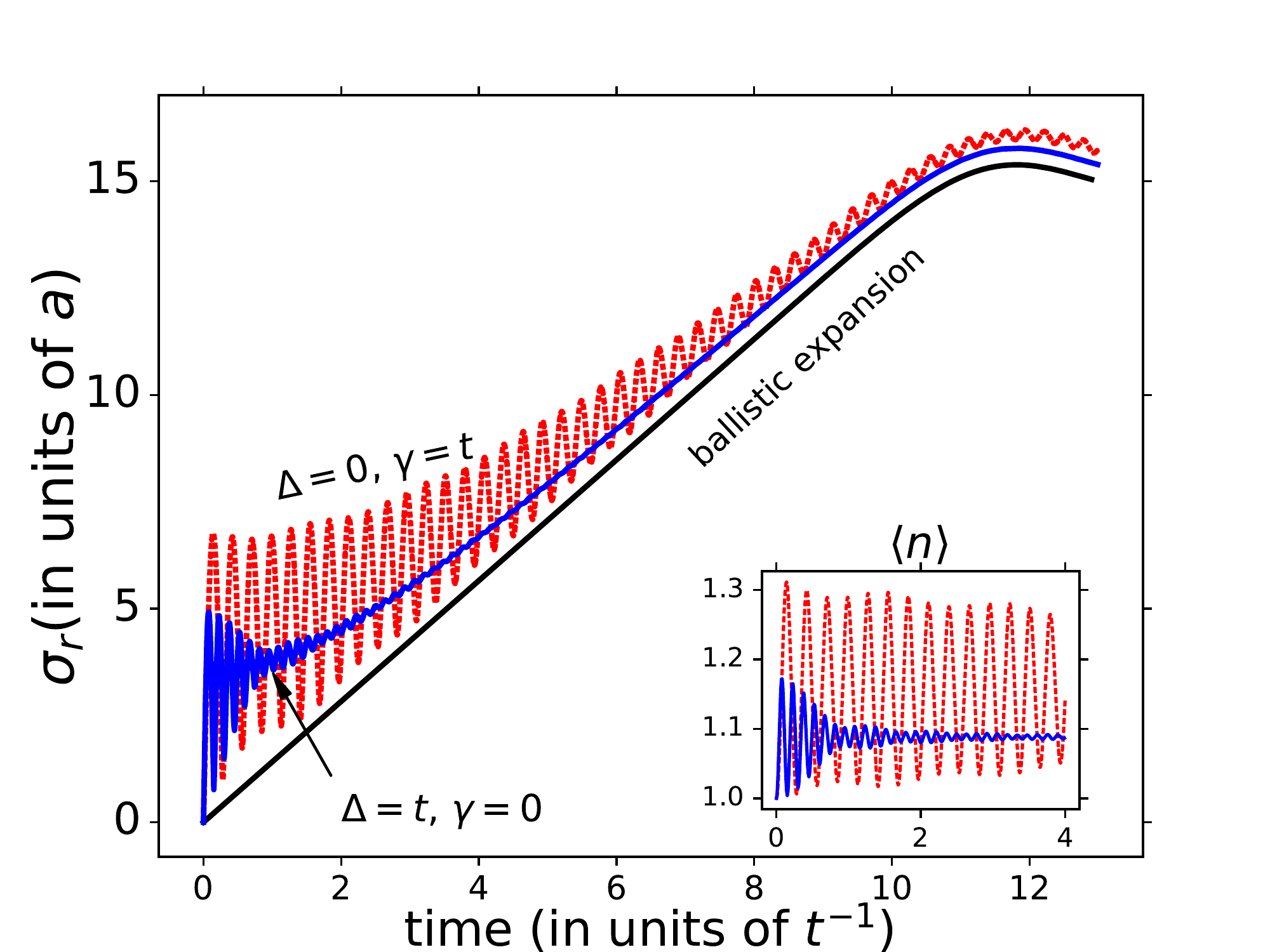} 
    \caption{Time dependence of the standard deviation (in units of lattice constant) of the wave packet for a particle initially placed in a single site of a one-dimensional ideal lattice. The solid black line represents the ballistic expansion governed by the Hamiltonian (\ref{model}) with $v=0$, $\Delta = 0$ and $\gamma = 0$. 
 For the dotted red curve $\Delta = 0$ and $\gamma = t$ while for the blue solid curve $\Delta = t$ and $\gamma = 0$. For all of these calculations, $\Delta \varepsilon = 20/t$. The inset shows the average number of particles $\langle n \rangle$ as a function of time for $\Delta = 0$ and $\gamma = t$ (dotted red curve) and $\Delta = t$ and $\gamma = 0$ (blue solid curve).
}
    \label{fig:sigma_vs_time_delta-and-gamma}
\end{figure}

Since the $c^\dagger_i c^\dagger_j$ term generates pairs of particles in adjacent sites, it is important to consider the role of inter-site interactions $v$. Such interactions appear in extended Hubbard models, leading to non-trivial properties of the lattice systems \cite{extended-Hubbard_1,extended-Hubbard_2,Vargas-Krems} and inducing correlations in quantum walks \cite{Tirtta-Krems}. Here, they are transient as the mutliple-particle subspaces are populated only virtually.  The inset of Figure 1 illustrates that repulsive interactions stabilize the oscillations at long times, while the short-time dynamics appears  to be largely unaffected by the density-density interactions. 


The $\Delta \neq 0$ term couples the subspaces with the odd number of particles. Thus, the state of a single particle is coupled to a state of three-particles, but not to the state of two particles or the vacuum state. By contrast, the $\gamma \neq 0$ term couples subspaces differing in the number of particles by one. To illustrate the effect of such couplings on the dynamics of quantum walks, we compare two models: (i) $\Delta = 0, \gamma = t$ and (ii) $\Delta = t, \gamma = 0$. The results shown in Figure 3 illustrate that the couplings in case (i) have a much stronger effect, leading to larger amplitudes of the oscillations and the persistence of the oscillations for much longer time.

\subsection{Disordered 1D lattices}

We next consider disordered 1D lattices. The disorder is generated by randomizing the on-site energy $\varepsilon_i$ by drawing the random values from a uniform distribution $[w/2, w/2]$, where $w$ quantifies the strength of disorder. Non-interacting particles are exponentially localized in 1D disordered systems \cite{anderson}. Our goal is to explore the role of the $\Delta \neq 0$ interactions on the localization.

In all of the disordered models we consider $\gamma = 0$, $\Delta/t \le 1$ and $\Delta \varepsilon/t = 20 $. 
Notice that for the ideal lattice with $\Delta / t = 1$ illustrated in Figure 1, this value of $\Delta \varepsilon$ ensures that the average number of particles $\langle n \rangle < 1.2$ at all times.
The three-particle sub-space is thus far off-resonant and contributes to the dynamics perturbatively.

 Figure 4 (upper panel) shows the average lattice population distributions illustrating the localization. To obtain these distributions, we place a particle in a single lattice site, propagate the wave packet to long time and average the resulting probability distribution over 100 random instances of disorder. We have verified that this number of disorder realizations ensures converged results. The averaging removes the time-dependence in the long-time limit. The results show that the term $\hat V_{\rm nc}$ induces non-exponential wings of the distribution, which rise with the magnitude of $\Delta$. To illustrate the quantitative contribution of these wings, we compute the inverse participation ratio (IPR) defined as
\begin{eqnarray}
I(t) = \sum_i\left( \frac{ |\psi_i(t)|^2}{\sum_i |\psi_i(t)|^2}\right)^2,
\label{ipr}
\end{eqnarray}
where $ |\psi_i(t)|^2$ is the probability of the population of lattice site $i$ at time $t$.
The value of the IPR ranges from $1/N$ for the state completely delocalized over the lattice with $N$ sites to 1 for the state localized in a single lattice site. 
We find that the couplings with $\Delta/t = 1$ decrease the IPR, indicating decrease of localization. 
Surprisingly, the effect of these couplings increases with increasing disorder strength. 
This phenomenon is reminiscent of noise-induced delocalization \cite{Mohseni-Aspuru,Rebentrost-Aspuru,noise_1,noise_2}. Here, the variation of on-site energy due to disorder brings the energy of the different particle-number states for random lattice sites closer together, thereby enhancing the effect of the couplings induced by $\hat V_{\rm nc}$. With increasing disorder strength $w$, the probability of the different number states becoming closer in energy increases, leading to more and stronger high-order hopping terms, thereby decreasing localization more significantly. Note that this result applies only in the limit $\Delta \ll \Delta \varepsilon$, i.e. in the limit where the number-changing interactions are much weaker than the energy separation between the number subspaces.

\begin{figure}[h!]
  \centering
    \includegraphics[width=0.5\textwidth]{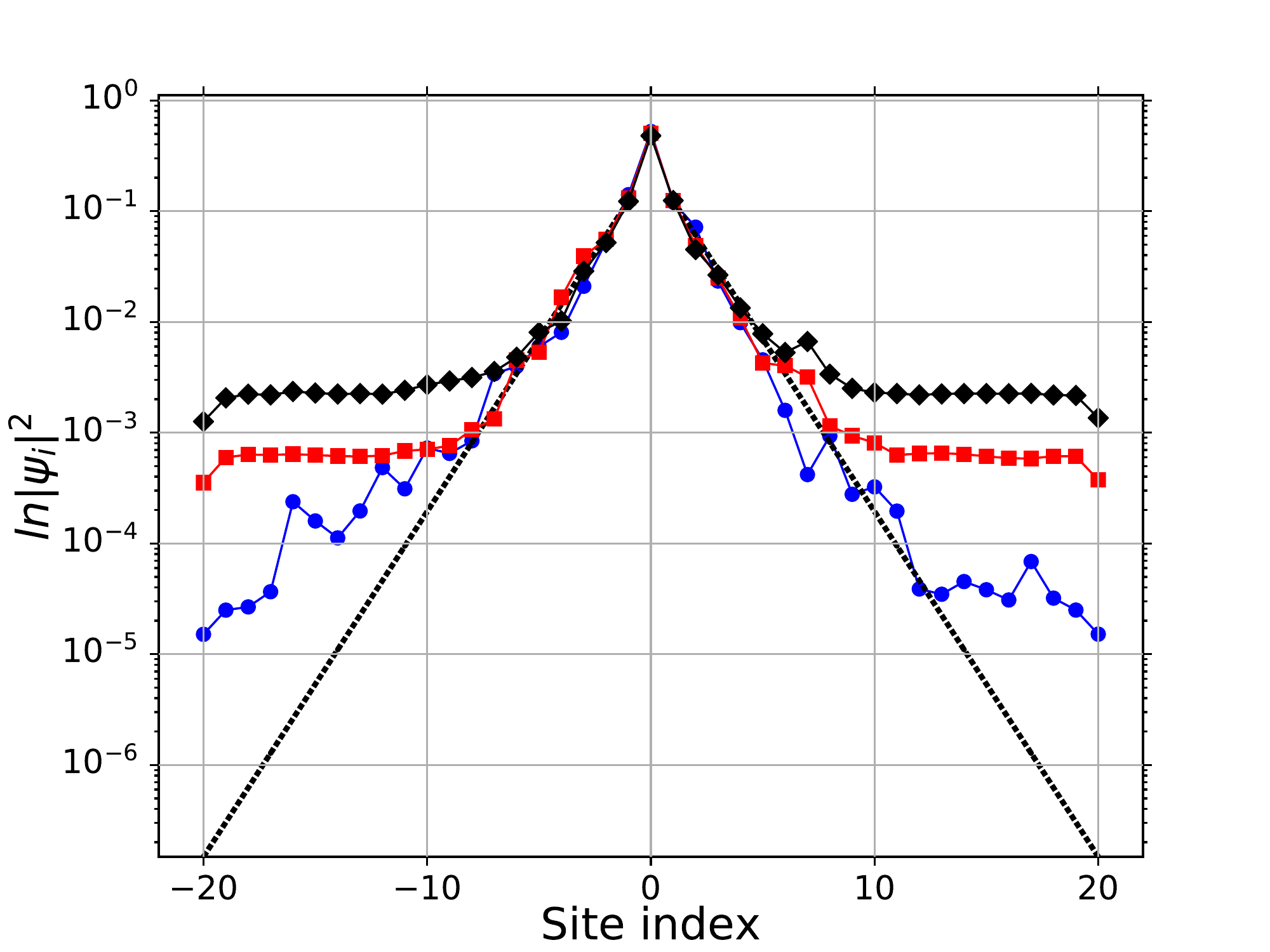}  \\
        \includegraphics[width=0.5\textwidth]{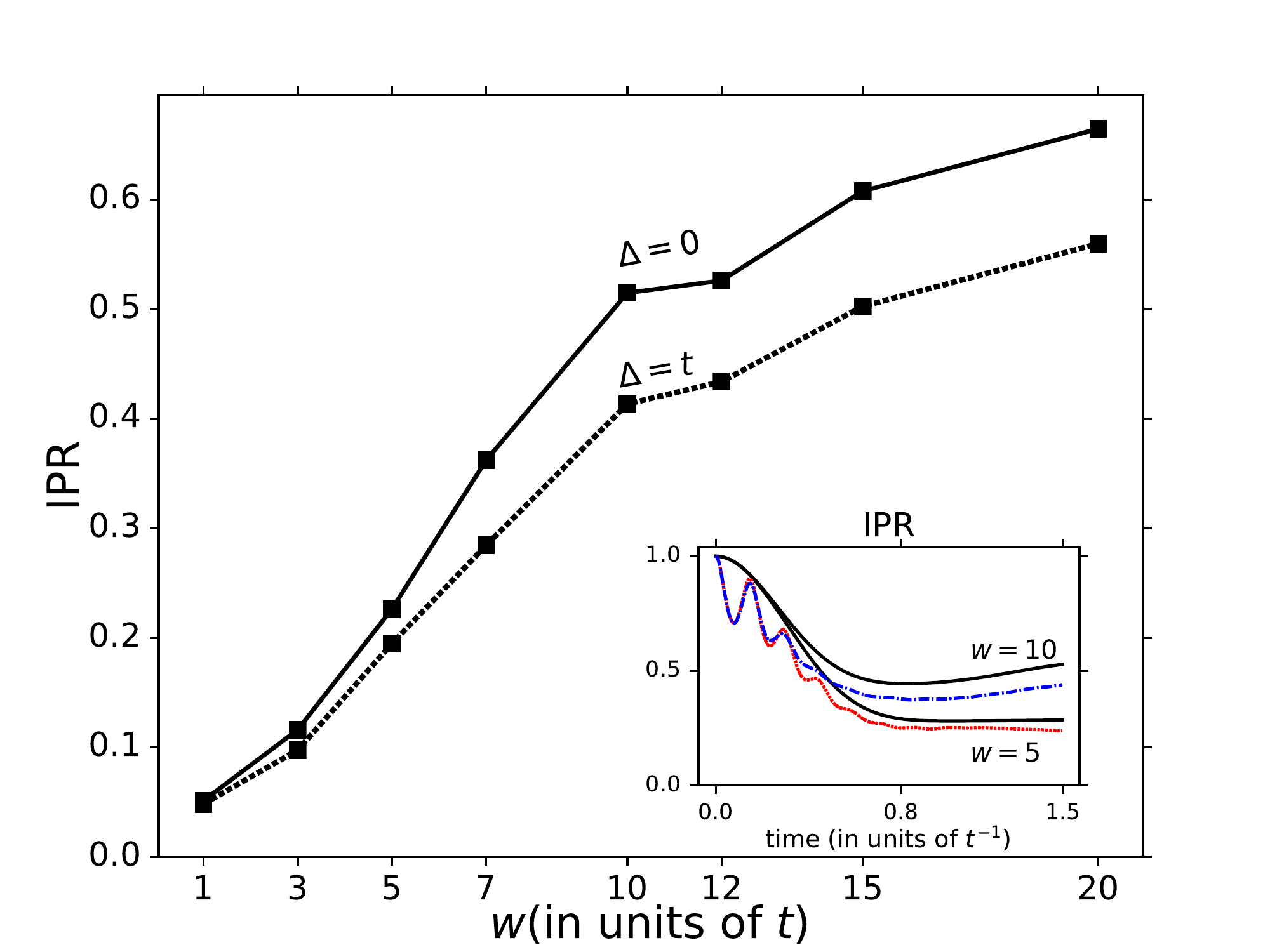}
    \caption{Upper panel: The logarithm of the particle probability distributions in a disordered 1D lattice with $41$ sites: diamonds -- $\Delta/t = 1$, squares -- $\Delta/t = 1/2$, circles -- $\Delta/t = 1/10$.  The results are averaged over 100 realizations of disorder and are time-independent. 
 The dashed line is an exponential fit to the  $\Delta/t = 1/10$ results.
 Lower panel: the long-time limit of the IPR defined in Eq. (\ref{ipr}) averaged over 100 instances of disorder as a function of the disorder strength $w$: solid line -- $\Delta = 0$, dashed line -- $\Delta/t =1$. The inset shows the IPR averaged over 100 realizations of disorder for two disorder strengths $w=\{5/t,10/t\}$ as functions of time: the solid black curves -- $\Delta = 0$; the dotted and dot-dashed curves -- $\Delta = t$. 
 }
    \label{fig:PProb_vs_sites}
\end{figure}

\begin{figure}[h!]
  \centering
 \includegraphics[width=0.4\textwidth]{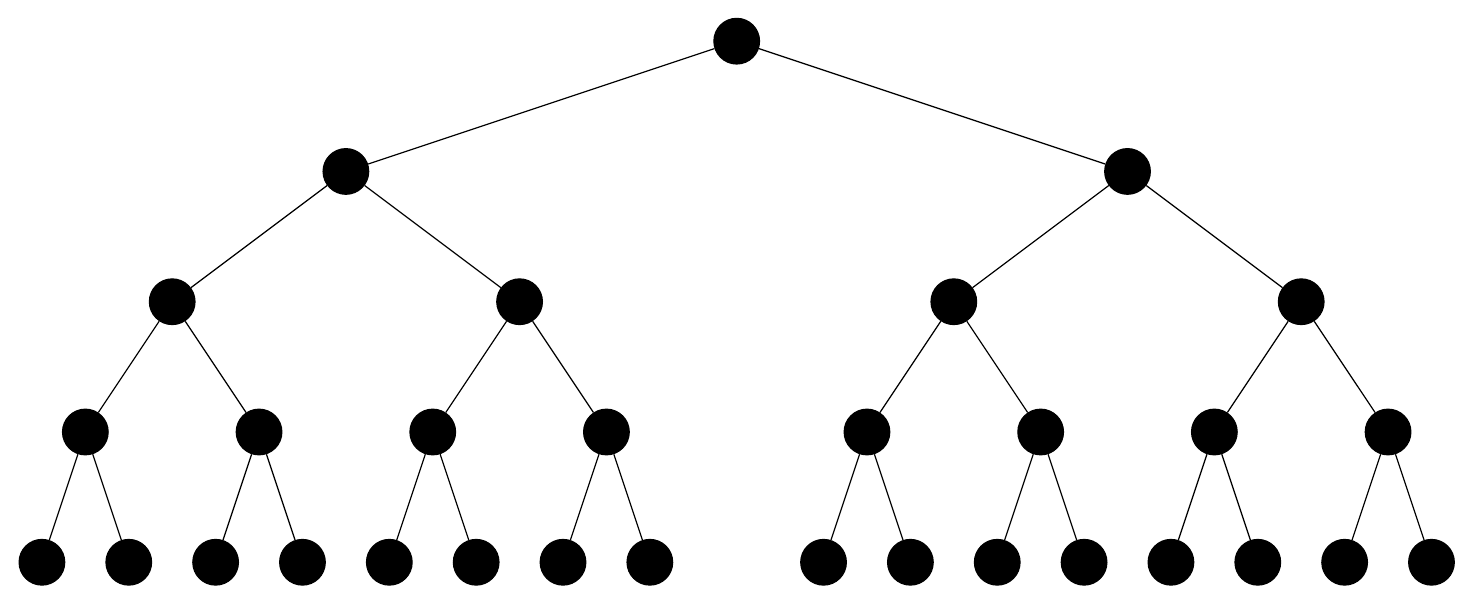} 
    \caption{ 
   Schematic diagram of an ideal binary tree with depth-$5$ ($\mathcal{G}5$). 
    }
    \label{fig:G5}
\end{figure}

\begin{figure}[h!]
  \centering
    \includegraphics[width=0.4\textwidth]{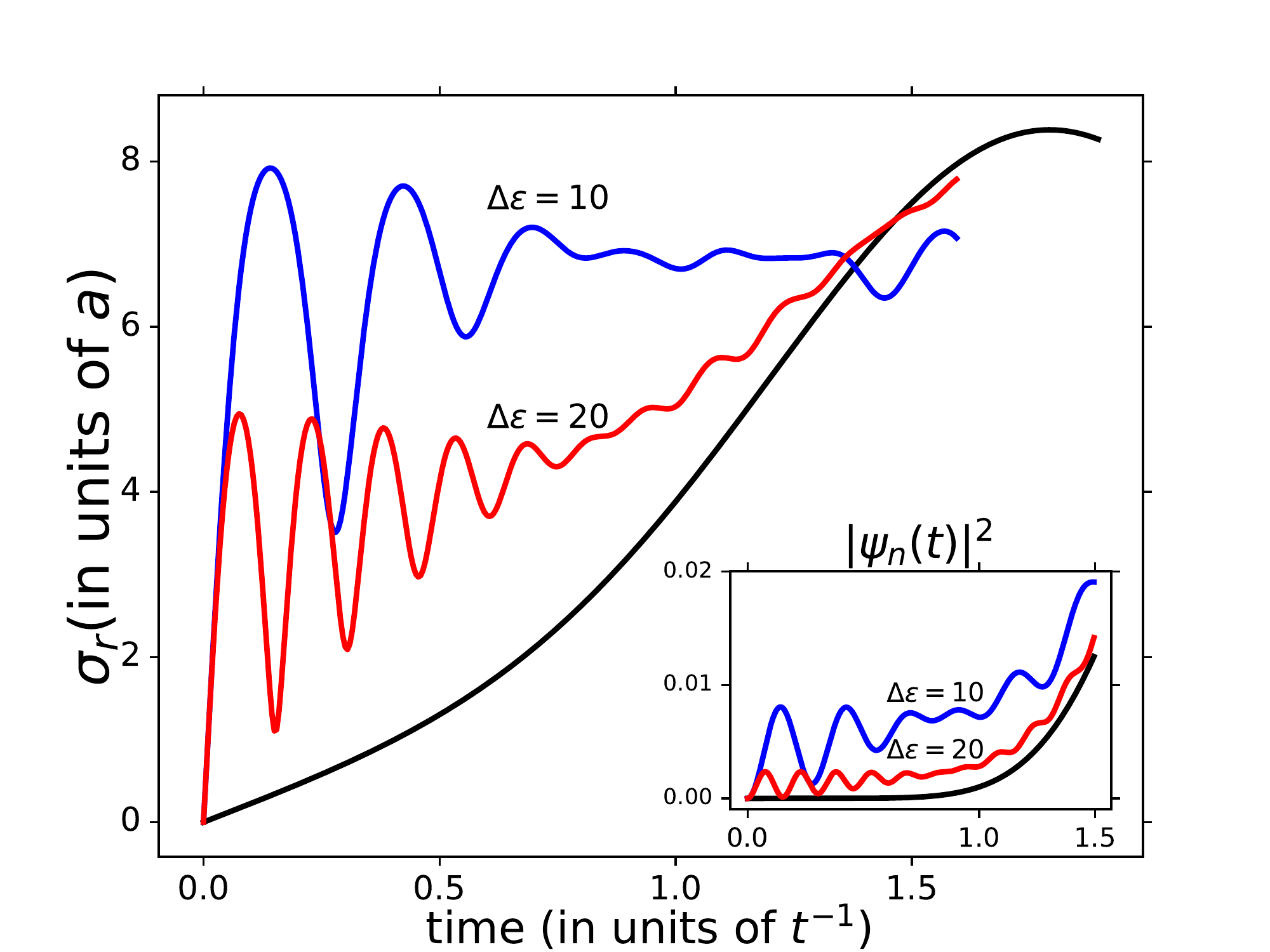} \\
     \includegraphics[width=0.4\textwidth]{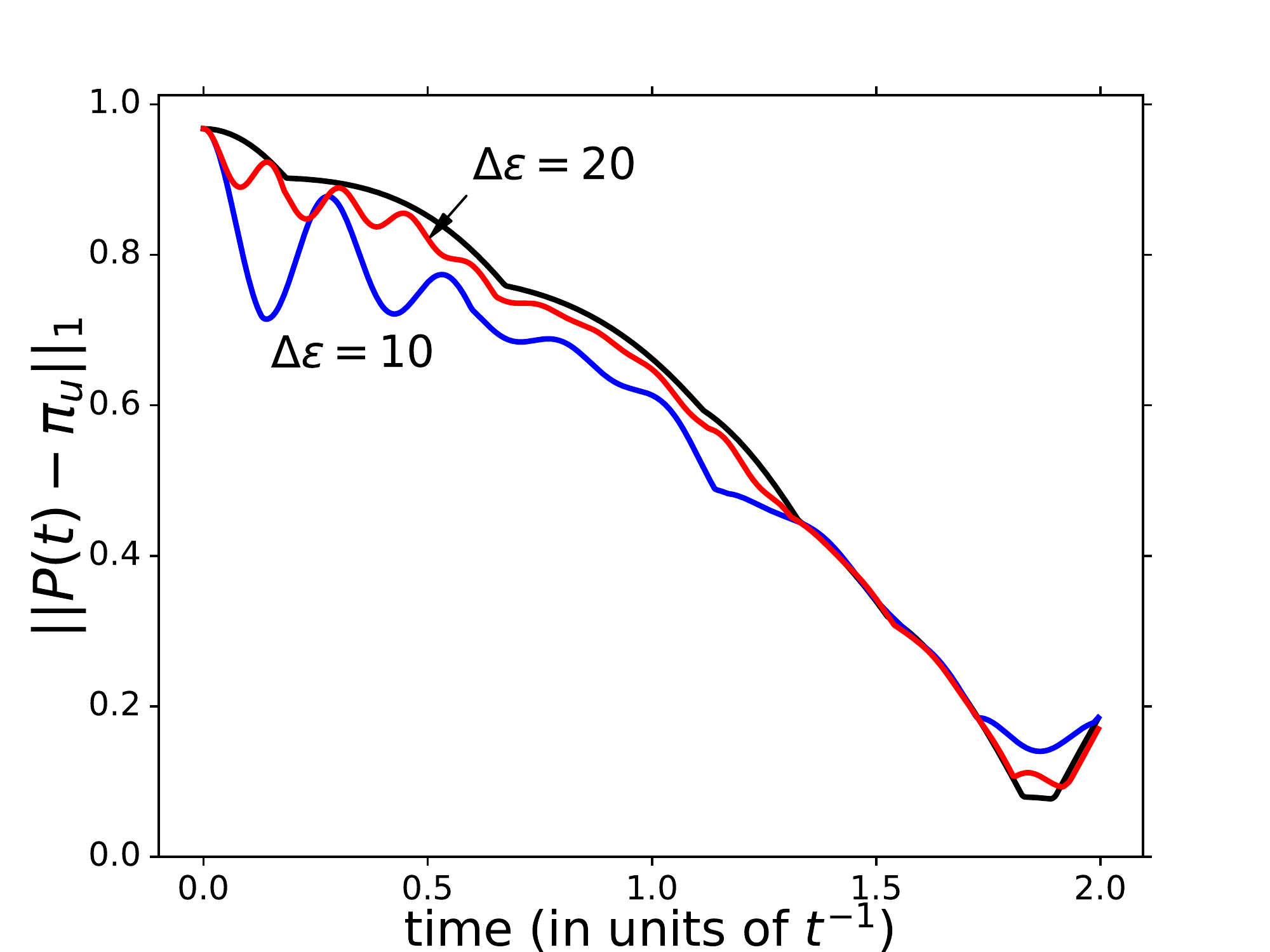}
    \caption{ 
      Upper panel: The growth of the wave packet for a single particle placed at the root of the tree ($\mathcal{G}5$): black line -- $\Delta = 0$; the oscillating curves -- $\Delta/t = 1$ with $\Delta \varepsilon = 10/t$ (blue)  and $\Delta \varepsilon = 20/t$ (red).
The inset shows the probability of reaching the last node (in layer 5) of the binary tree as a function of time. The curves are color-coded in the same ways as in the main plot. 
    Lower panel: The $L1$ norm between the probability distribution of the wave packet and the uniform distribution as a function of time. The uniform distribution is defined by the value $1/{\rm dim}(\mathcal{G}5)$ for each node. The solid black curve is for $\Delta = 0$. The oscillating curves are for $\Delta=t$ with $\Delta \varepsilon = 10/t$ (blue) or $\Delta \varepsilon = 20/t$ (red). For all curves in both figures $\gamma = 0$.
    }
    \label{fig:G5}
\end{figure}

\begin{figure}[h!]
  \centering
    \includegraphics[width=0.4\textwidth]{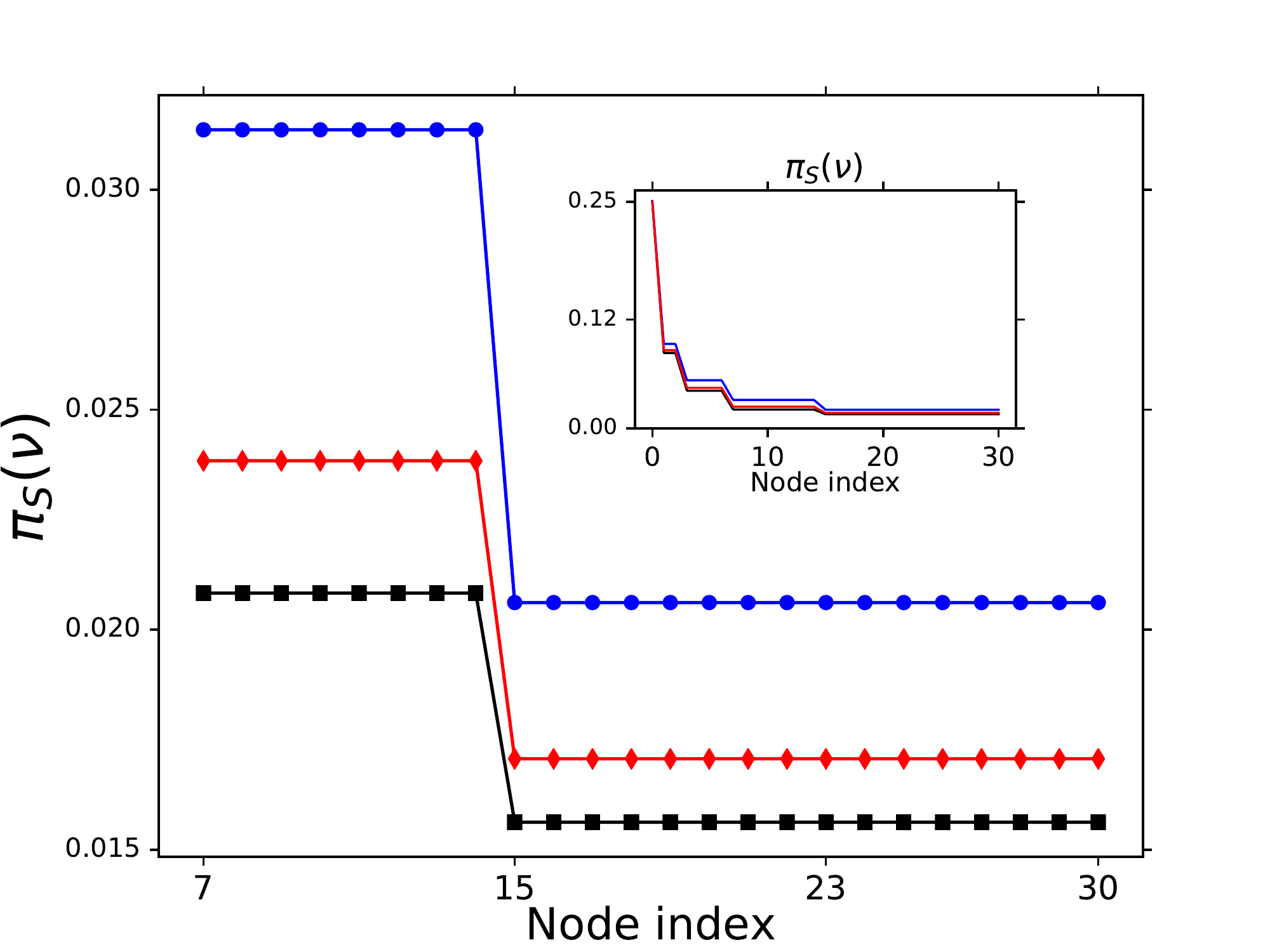}
    \caption{ 
    	The squares represent the stationary distribution $\pi_S(\nu)$ for a quantum walk on the $\mathcal{G}5$ tree with $\Delta = 0$. The circles and diamonds are the the stationary distributions $\pi_S(\nu)$ for quantum walks with $\Delta/t = 1$ and $\Delta \varepsilon = 10/t$  and $\Delta \varepsilon = 20/t$, respectively. 
    }
    \label{fig:G5}
\end{figure}

\subsection{Binary trees}

We next consider quantum walks in binary trees. A binary tree is characterized by the number of layers and the connectivity of the lattice sites. Here, we consider the binary tree ${\cal G}5$ with five layers schematically depicted in Figure 5.  The model (\ref{model}) adapted to binary trees becomes
\begin{eqnarray}
\mathcal{H} &=& \sum_{i=1}^{2^g-1} (\Delta \varepsilon + \varepsilon_i) \hat{c}^\dagger_i\hat{c}_i +  \sum_{i=1}^{2^g-1} \sum_{j=1}^{2^g-1} t_{ij}\left (\hat{c}^\dagger_i\hat{c}_j + \hat{c}^\dagger_j\hat{c}_i \right) \nonumber \\
 &&+ \sum_{i=1}^{2^g-1} \sum_{j=1}^{2^g-1} \Delta \left (\hat{c}^\dagger_i\hat{c}^\dagger_j + \hat{c}_j\hat{c}_i \right)
\label{binary-tree-model}
\end{eqnarray}
We set $\gamma = 0$ for all binary tree calculations. 
Each node of the binary tree is connected to three nodes: its father, the left child $2i$ and the right child $2i+1$, so
\begin{eqnarray}
\mathcal{H} &=& \sum_{i=1}^{2^g-1} (\Delta \varepsilon + \epsilon_i) \hat{c}^\dagger_i\hat{c}_i +  t \sum_{i=1}^{2^g-1}  \left (\hat{c}^\dagger_i\hat{c}_{2i} + \hat{c}^\dagger_i\hat{c}_{2i+1}+ h.c. \right)  \nonumber\\ &&+ \hat V_{\rm nc, tree}
\end{eqnarray}
where
\begin{eqnarray}
\hat V_{\rm nc, tree} =  \Delta \sum_{i=1}^{2^g-1} \left (\hat{c}^\dagger_i\hat{c}^\dagger_{2i} + \hat{c}^\dagger_i\hat{c}^\dagger_{2i + 1} + \hat{c}_i\hat{c}_{2i}  + \hat{c}_i\hat{c}_{2i + 1} + h.c.\right) \nonumber\\
\end{eqnarray}

We describe the spread of the quantum wave packets in such trees as 
\begin{eqnarray}
\sigma(t) = \sqrt{\langle \nu^2\rangle- \langle \nu \rangle^2} = \sqrt{\sum_{\nu =1}^{2^g-1} \nu^2 p_\nu(t) - \left ( \nu p_\nu(t)\right )^2 }
\end{eqnarray}
We consider quantum walks starting at the root of a graph and compare the dynamics in models with $\Delta = 0$ and $\Delta \neq 0$.

Figure 6 (upper panel) shows that the couplings $\hat V_{\rm nc, tree}$ accelerate quantum walks on the tree. To quantify the effect of $\Delta \neq 0$ on quantum walks, we compute the mixing time defined as 
\begin{eqnarray}
M_\epsilon = \min\left \{ T | \forall \; t \geq T :\left \| P(t) - \pi \right \|_1 \leq \epsilon \right \}
\label{eqn:mixtime}
\end{eqnarray}
where $P(t)$ is the probability distribution at time $t$, $\pi$ is a distribution that the quantum system is expected to approach, and $\left \| \cdot \right \|_1$  is the $L1$ norm. 

We consider two distributions $\pi$: the uniform distribution  $\pi_U$ and the stationary distribution $\pi_S$. 
The uniform distribution is characterized by the same value of probability for each node. 
The stationary distribution is defined by the following values of the probability for node~$\nu$ 
\begin{eqnarray}
\pi_S(\nu) = \lim_{T\to\infty}\bar{p}_\nu(T),
\label{eqn:limdens}
\end{eqnarray}
where
$\bar{p}_\nu(T)$ is the time average of the probability of populating node $\nu$, defined as 
\begin{eqnarray}
 \bar{p}_\nu(T) &=& \frac{1}{T}\sum_{t=0}^{T-1}p_\nu(t) = \frac{1}{T}\sum_{t=0}^{T-1}\left | \langle \nu | \Psi(t) \rangle \right|^2 \\
 &=& \frac{1}{T}\sum_{t=0}^{T-1} \langle \nu | \Psi(t) \rangle\langle  \Psi(t) | \nu \rangle \nonumber \\
&=& { \frac{1}{T}\sum_{t=0}^{T-1} }  \left \{ \sum_\lambda e^{iE_\lambda t}\langle \nu |\lambda\rangle \langle \lambda | \Psi(0) \rangle  \right \}  \nonumber \\
 &\times &\left \{  \sum_{\lambda'} e^{iE_{\lambda'} t}\langle\Psi(0) |\lambda'\rangle \langle \lambda' |\nu \rangle \right \} 
\end{eqnarray}
Here, we set $\hbar = 1$. Defining $\langle \lambda | \Psi(0) \rangle$  as $c_{n_0}^\lambda$,
 \begin{eqnarray}
 \bar{p}_\nu(T) &=& \frac{1}{T}\sum_{t=0}^{T-1} \left \{ \sum_\lambda e^{iE_\lambda t}c_{n_0}^\lambda\langle \nu |\lambda\rangle  \right \}
\left \{  \sum_{\lambda'} e^{iE_{\lambda'} t}c_{n_0}^{\lambda' *} \langle \lambda' |\nu \rangle \right \} \nonumber \\
&=& \frac{1}{T}\sum_{t=0}^{T-1}\sum_\lambda |c_{n_0}^\lambda |^2 \;  |\langle \nu |\lambda\rangle|^2 \nonumber\\
&&+ \frac{1}{T}\sum_{t=0}^{T-1} \sum_{\lambda,\lambda'}\left (c_{n_0}^\lambda  c_{n_0}^{\lambda' *} e^{i(E_\lambda - E_{\lambda'})t} \langle \nu |\lambda\rangle \langle \lambda' |\nu \rangle \right  ).
\end{eqnarray}
In the limit of long time $T \to \infty$ the imaginary part of  $\bar{p}_\nu(T)$ tends to zero. The factor $\frac{1}{T}$ in the real part of $ \bar{p}_\nu(T)$ cancels because  $\sum_{t=0}^{T-1} e^{i(E_\lambda - E_{\lambda})t} =  T$.  We can thus rewrite $\pi_S(\nu)$ as
 \begin{eqnarray}
\pi_S(\nu) &=&\sum_\lambda |c_{n_0}^\lambda |^2 \;  |\langle \nu |\lambda\rangle|^2. 
 \label{eqn:limdens_2}
\end{eqnarray}
From Eq. (\ref{eqn:limdens_2}) we observe that $\pi_S(\nu)$ depends on the initial condition ($|\Psi(t=0) \rangle$). 

Figure 6 (lower panel) illustrates the effect of the couplings $\hat V_{\rm nc, tree}$  on the speed of approaching the uniform distribution $\pi_u(\nu)$ and Figure 7 the effect of the couplings $\hat V_{\rm nc, tree}$ on the stationary distribution $\pi_S(\nu)$.  The approach to the uniform distribution is accelerated by the $\hat V_{\rm nc, tree}$ terms at short times. As can be seen from Figure 6, the couplings $\hat V_{\rm nc, tree}$ enhance the stationary distribution, illustrating that the graph is explored more efficiently by the dynamics with the $\hat V_{\rm nc, tree}$ couplings.

\begin{figure}[h!]
  \centering
    \includegraphics[width=0.25\textwidth]{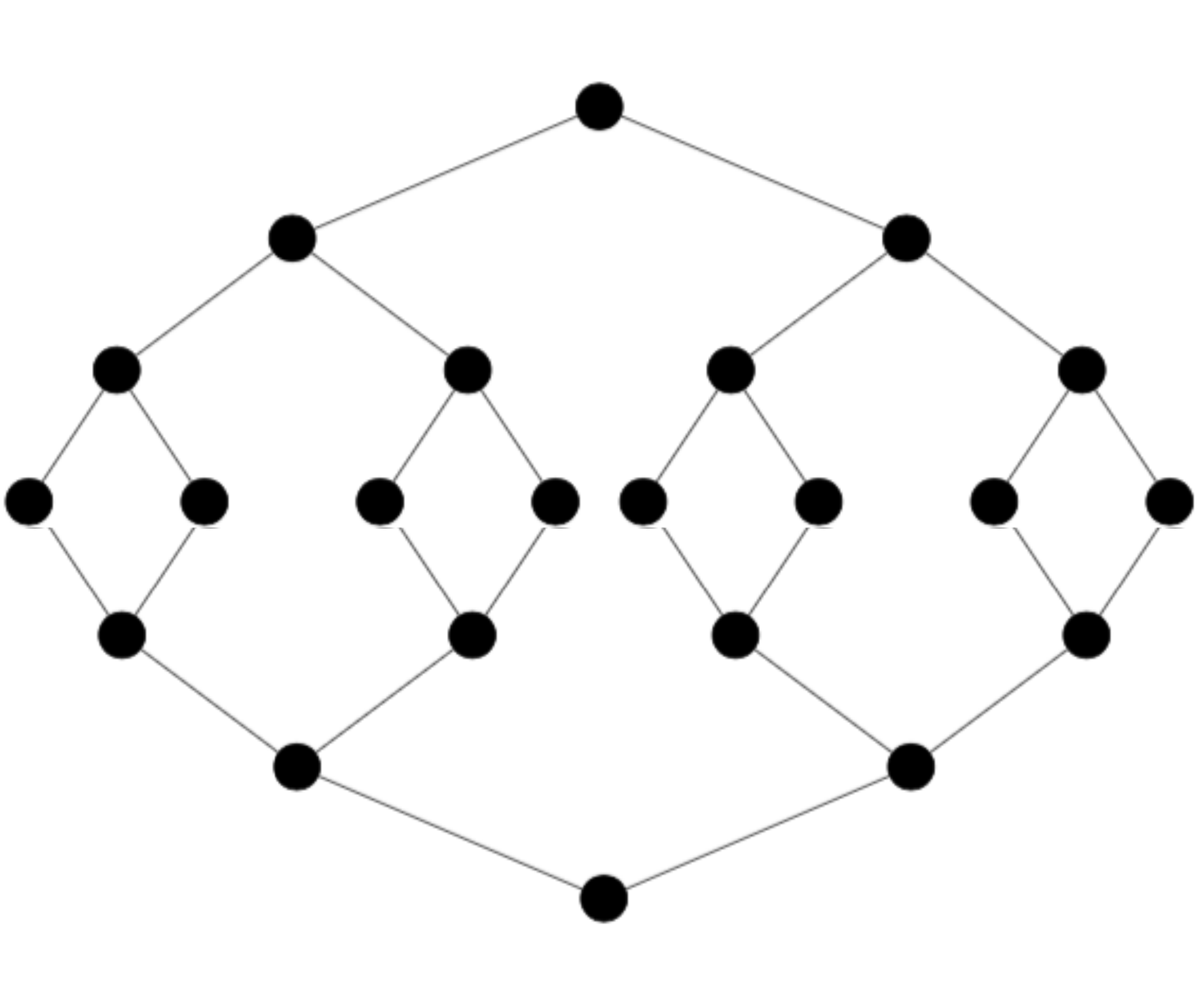} 
        \caption{ 
     Schematic diagram of an ideal glued binary trees with depth-$4$ ($\mathcal{GBT}4$).  
  }
    \label{fig:G5}
\end{figure}

\begin{figure}[h]
  \centering
    \includegraphics[width=0.5\textwidth]{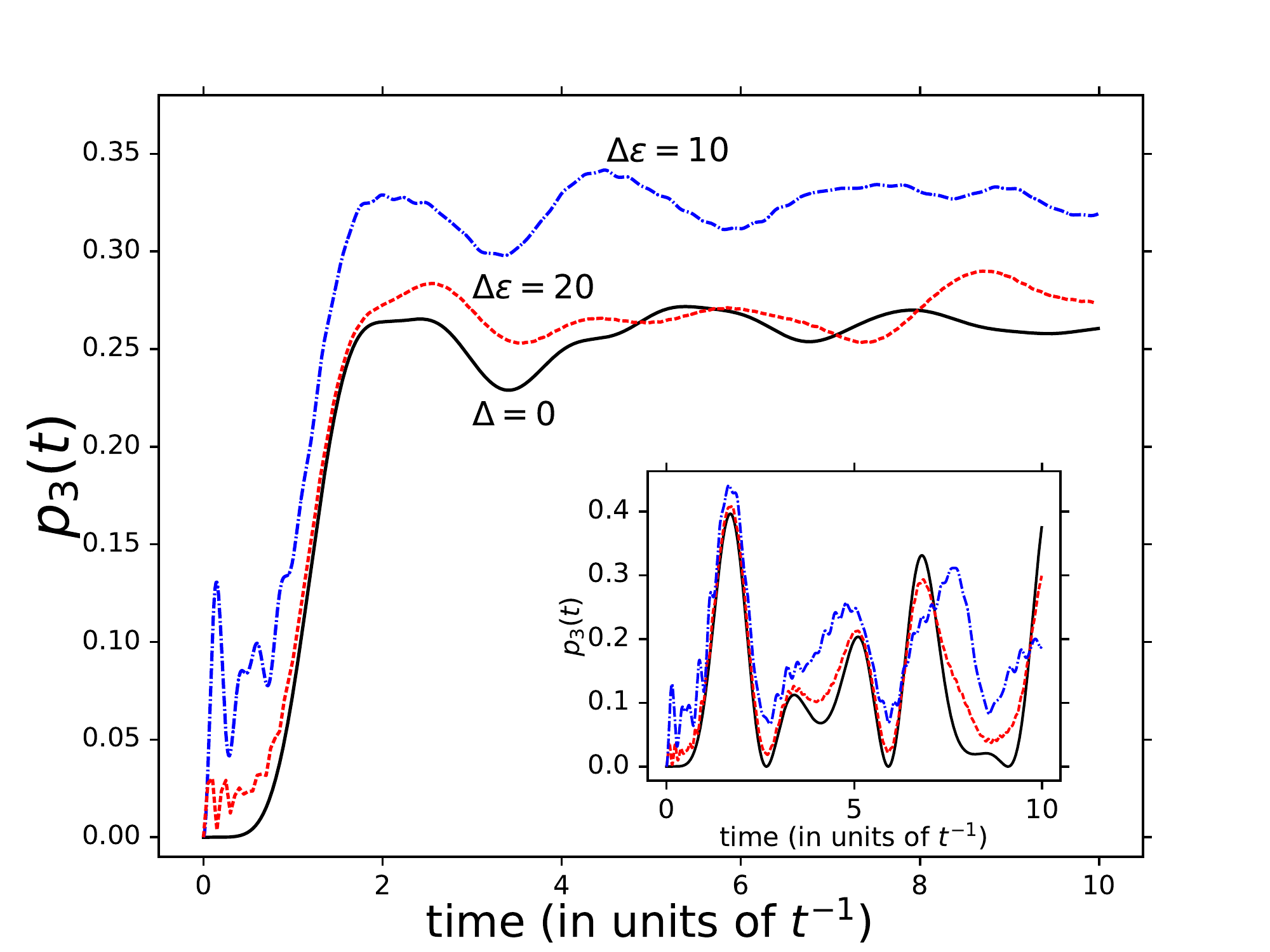}\\
      \includegraphics[width=0.5\textwidth]{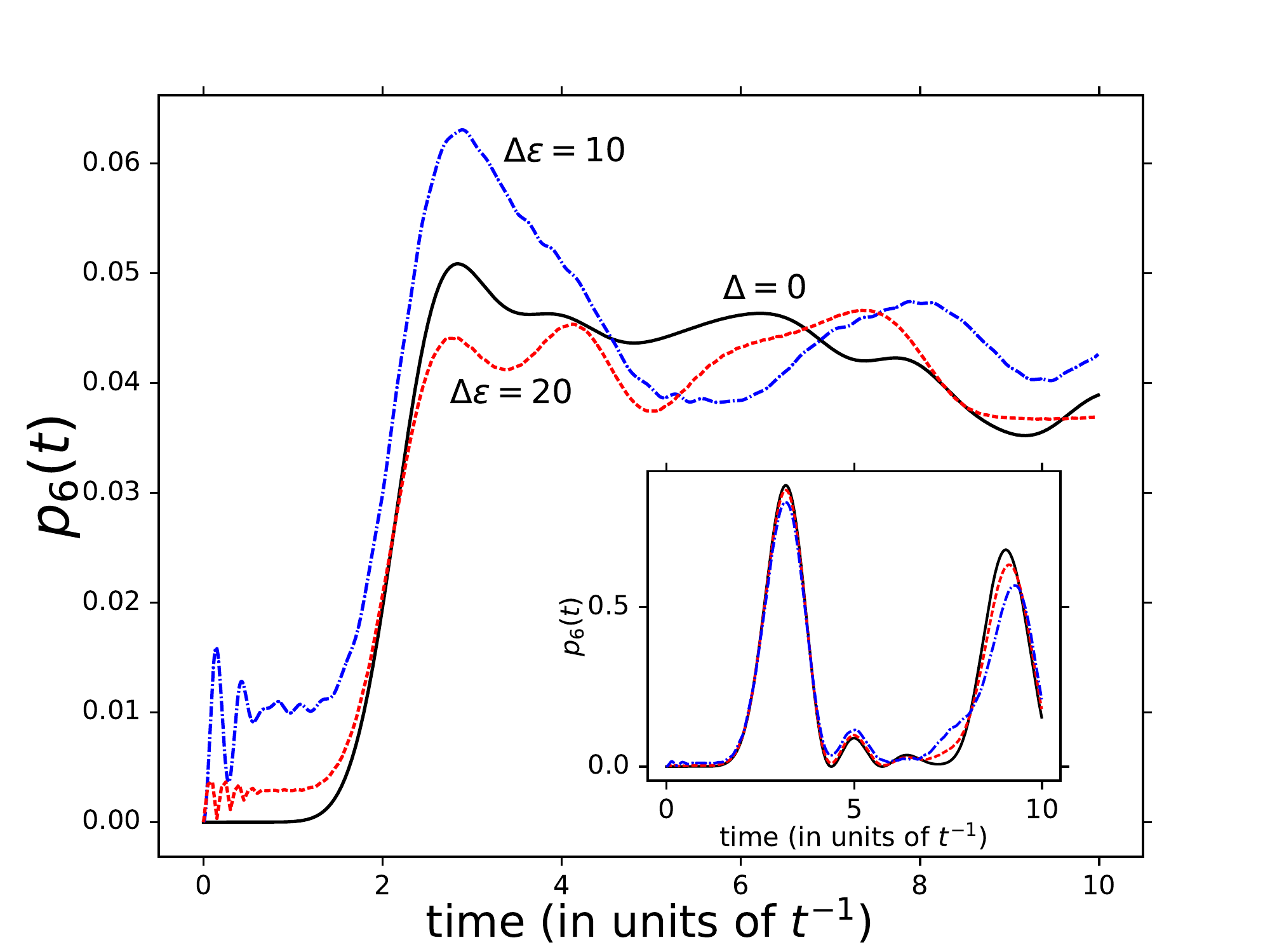}
    \caption{ 
    Average particle probability distribution in a disordered $\mathcal{GBT}4$ graph with strength $w = 5/t$. 
    The wave packet for a single particle is initially placed in the head node of a $\mathcal{GBT}4$ graph shown in Figure 8.
    The upper panel displays the probability (\ref{summed-probability}) summed over all nodes of layer $j=3$;  the lower panel shows $p_j(t)$ for the bottom node $j=6$.
    Insets: The particle probability distributions for an ideal $\mathcal{GBT}4$ graph with $\Delta \epsilon = 10/t$ (blue dot-dashed) and $\Delta \epsilon = 20/t$ (red dashed). The broken curves show the results obtained with $\Delta/t = 1$ and the full black lines -- $\Delta/t = 0$.
    }
    \label{fig:9}
\end{figure}

\begin{figure}[h]
\hspace{-1cm}
    \includegraphics[width=0.55\textwidth]{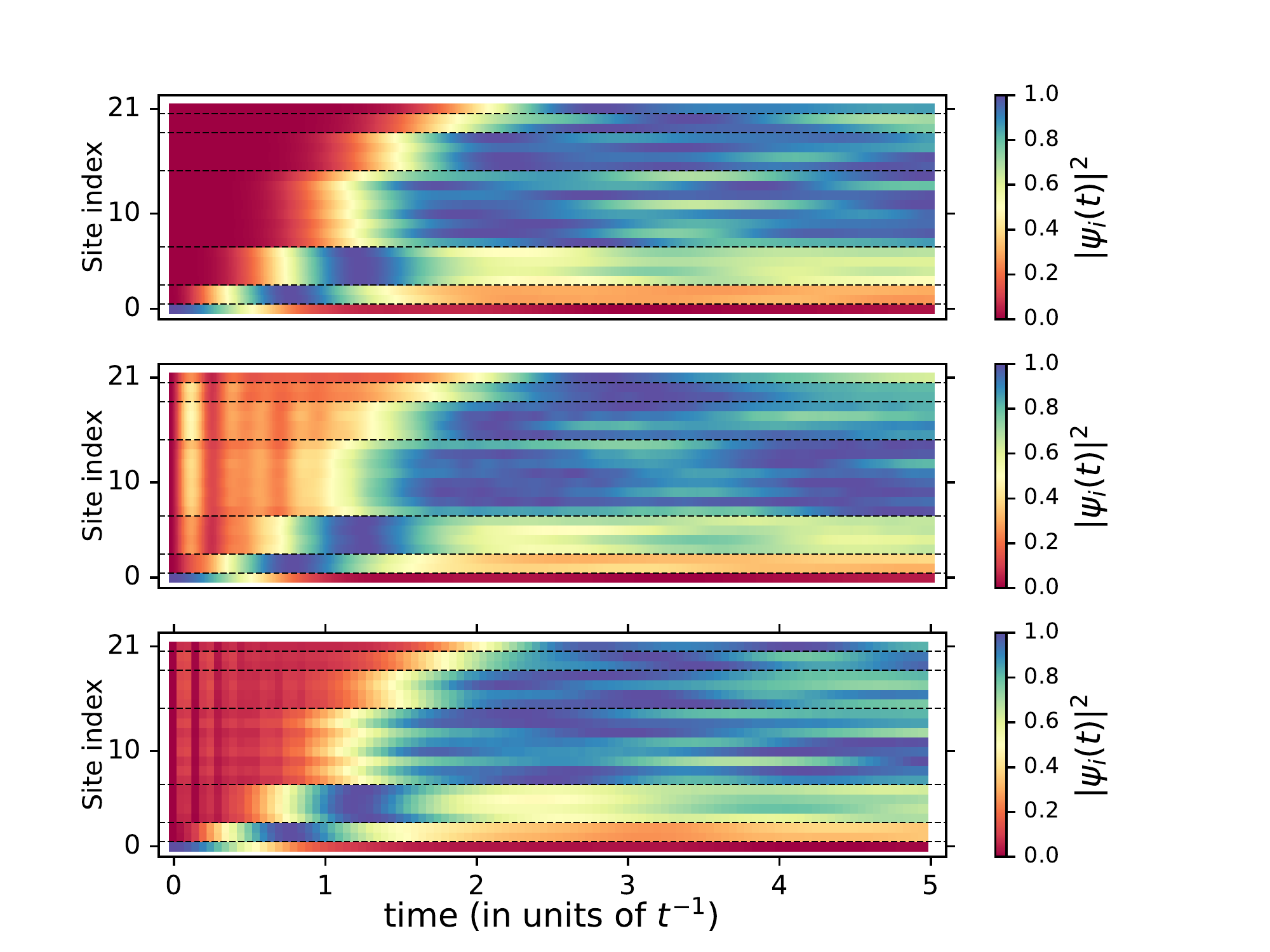} 
    \caption{ 
    Average particle probability distribution in a disordered $\mathcal{GBT}4$ graph: upper panel $-$ $ \Delta/t = 0$, middle panel $-$ $\Delta \epsilon= 10/t$ and lower panel $-$ $ \Delta \epsilon= 20/t$. For all panels we consider 100 realizations of disorder with a strength of $w = 5/t$ and $\gamma = 0$.
    The wave packet for a particle is initially placed in the head node of a $\mathcal{GBT}4$ graph.
    }
    \label{fig:8}
\end{figure}

If two binary trees of Figure 5 are joined together as shown in Figure 8, one obtains a glued binary tree. Transport through glued binary trees represents an important class of problems 
\cite{gluedtrees,gluedtrees2}. Of particular interest is the probability of transfer from the head node to the bottom node in disordered glued trees. 
Studies of such processes have been used to understand the consequences of quantum localization for the application of quantum walks for quantum computing and quantum communcation algorithms \cite{gluedtrees,gluedtrees2}.

To study quantum walks in a glued binary tree, we use the model (\ref{binary-tree-model}) but  with the summation index adapted to the tree shown in Figure 8. 
Figures 9 and 10 illustrate the effect of the particle-number fluctuations on quantum walks through a glued binary tree. 
 We first consider quantum walks in an ideal glued tree. 
 The results shown in the insets of Figure 9 are for a single particle placed at zero time in the head node of the glued tree depicted in Figure 8. 
 The upper panel of Figure 9 shows the probability 
\begin{eqnarray}
p_{j}(t) = \sum_{j=0}^{2^d}| \langle j | \psi(t) \rangle|^2
\label{summed-probability}
\end{eqnarray}
summer over all nodes of depth level $j = 3$. 
The lower panel is the probability of particle density transfer between the two ends of the glued tree. 
 Interestingly, while the $\Delta \neq 0$ interactions affect the population of the $j=3$ level, 
  we observe little effect of these interactions on the head-to-bottom transfer of the particle for times $< 10~t^{-1}$ (see the inset of the lower panel of Figure 9).

In contrast, the same interactions have a much stronger effect on the head-to-bottom transfer through a disordered tree. 
 As can be seen from Figure 9, the $\Delta \neq 0$ interactions 
accelerate the efficiency of particle transfer through the disordered tree, especially at short times by inducing oscillations as in the case of an open-ended binary tree discussed above.
Figure 10 shows that these oscillations survive averaging over 100 disorder realizations. 
We set the disorder strength to $w = 5/t$ for this calculation. 

While the methodology used here limits the size of the glued tree to seven levels, our results indicate that the localization of quantum particles in disordered glued trees must be affected by the 
couplings between particle number subspaces. It would be interesting to see if the head-to-bottom transfer remains insensitive to these interactions and how the localization length is affected by such interactions in larger trees. To treat such problems, it is necessary to develop approximate computation techniques for few-particle systems in glued trees.

\section{Conclusion}

In this work we consider coherent quantum dynamics governed by the lattice Hamiltonians with number non-conserving interactions in the few-body limit. 
We have illustrated that couplings between particle-number subspaces, even if much smaller than the energy separation between these subspaces, accelerate the dynamics of quantum walks in ideal lattices and binary trees and increase the localization length in disordered lattices. Effectively, these couplings provide new degrees of freedom, increasing the range of hopping due to virtual excitations and/or transient elimination of a single particle due to coupling to the vacuum state.  We have shown that the number-changing interactions decrease the mixing and hitting times for quantum walks on binary trees.  

Our results show that the inverse participation ratio in disordered one-dimensional lattices decreases in the presence of number-changing interactions, signalling decrease of localization. 
This effect increases with increasing disorder strength, leading to larger changes of the inverse participation ratio in lattices with stronger on-site disorder. This is a direct consequence of the disorder-induced broadening of the particle energy bands. This broadening brings different particle number subspaces closer in energy, increasing the effect of the number-changing couplings and, consequently, the effective range of particle hopping. 

Engineering lattice Hamiltonians to accelerate quantum dynamics has been of much recent interest due to potential applications in quantum computing and the study of the fundamental limits of the speed of correlation propagations in quantum many-body systems. Also of much interest is the localization dynamics of particles with long-range hopping in disordered lattices and graphs. 
Our work illustrates that models of the type (\ref{model}) can be used to study the effect of hopping range on Anderson localization and quantum walks spreading faster than ballistic expansion.

While non-interacting particles are known to be always localized in disordered 1D lattices, there is a localization - diffusion transition in 3D lattices \cite{root-skinner}. Our results indicate that the number-changing interactions must affect this transition. It would be interesting in future work to explore the quantitative effect of such interactions on the localization transition in 3D disordered lattices. It would also be interesting to explore the effect of such interactions on localization in 2D lattices. While non-interacting particles with short-range hopping are known to be always localized in 2D disordered latices, particle interactions may lead to delocalization.  Since the $\Delta \neq 0$ terms considered here create pairs of interacting particles in adjacent sites, these interactions may have non-trivial consequences on the localization in disordered 2D lattices.

\section*{Acknowledgments}

This work is supported by NSERC of Canada. 

\appendix
\section{Schrieffer--Wolff transformation of Hamiltonian (\ref{model})}
\label{ap:SW}
Here, we show that the particle-changing interactions in Eq. (\ref{pnnc}), if perturbative, modify the range of particle hopping. 
In particular, we show that, to leading order, the couplings (\ref{pnnc}) lead to next-nearest-neighbour hopping.

It is clear that the Hamiltonian (\ref{model}) is not block diagonal in the site representation basis due to couplings in Eq. (\ref{pnnc}). Using the Schrieffer--Wolff (SW) transformation, it is possible to block diagonalize this Hamiltonian.
We follow the notation in Ref. \cite{SW2}. The total Hamiltonian is defined as ${\cal \hat{H}} ={\cal \hat{H}}_0 +  {\cal \hat{H}}'$, where ${\cal \hat{H}}_0$ contains all the operators that commute with the particle number operator, and ${\cal \hat{H}}' = \hat V_{\rm nc}$.
The SW transformation assumes that the transformed Hamiltonian
\begin{eqnarray}
{\cal \tilde{H}} = e^{-S}{\cal \hat{H}}e^{S}
\end{eqnarray}
can be written as
\begin{eqnarray}
{\cal \tilde{H}} = H^{(0)} + H^{(1)} + H^{(2)} + \cdots,
\end{eqnarray}
with the different terms defined by the following matrix elements: 
\begin{eqnarray}
H^{(0)}_{mm'} &=& H^0_{mm'} \\
H^{(1)}_{mm'} &=& H'_{mm'} \\
H^{(2)}_{mm'} &=& \frac{1}{2} \sum_\ell \left [ \frac{1}{E_m - E_\ell} +  \frac{1}{E_m' - E_\ell}\right ] H'_{m\ell}H'_{\ell m'}. 
\quad \quad \label{eqn:SW_2}
\end{eqnarray}
Here, the indices $m$ and $m'$ refer to any one-particle state, while $\ell$ is an index of a three-particle state.  $H^0$ is the Hamiltonian (\ref{model})  with $\hat V_{\rm nc} = 0$. All matrix elements of $H^{(1)}$ are zero: $H^{(1)}_{mm'} = 0$. The first correction to ${\cal \tilde{H}}$ appears in $H^{(2)}$ whose matrix elements depend on $\hat V_{\rm nc}$. Here we only consider the case of a 1D lattice with $\gamma = 0$, $\Delta \neq 0$, and nearest-neighbour interactions. The matrix elements $H^{(2)}_{mm'}$ depend on the matrix elements of the $\Delta$--dependent term in $\hat V_{\rm nc}$,
\begin{eqnarray}
H'_{m\ell} &=& \langle m |\Delta \sum_{i} (\hat{c}^{\dagger}_{i}\hat{c}^\dagger_{i\pm 1} + \hat{c}_{i}\hat{c}_{\pm 1})  | abc\rangle = \Delta \sum_{i} \langle m | \hat{c}_{i}\hat{c}_{i\pm 1}  | abc\rangle \nonumber\\ 
&=&  \Delta \left [ \delta_{m,c}\left (\delta_{b\pm1,a} +\delta_{a\pm1,b}  \right ) + \delta_{m,b} \left( \delta_{c\pm 1, a} +\delta_{a\pm1,c}  \right ) \right. \nonumber \\
&&\left.+ \delta_{m,a}\left( \delta_{c\pm1,b} + \delta_{b\pm1,c}\right )\right ]\label{eqn:mm_1}
\end{eqnarray}
where $a$, $b$ and $c$ are the lattice indices of the three particles.
In the case of an ideal system, Eq. (\ref{eqn:SW_2}) can be rewritten as, 
\begin{eqnarray}
H^{(2)}_{mm'} &=& -\frac{1}{2\Delta \epsilon} \sum_\ell  H'_{m\ell}H'_{\ell m'} 
 \label{eqn:SW_2_ideal}
\end{eqnarray}
where the summation $\sum_\ell$ is over all possible combinations of lattice indices for three particles.
Inserting Eq. (\ref{eqn:mm_1}) into Eq. (\ref{eqn:SW_2}) for both  $H'_{m\ell}$ and $H'_{\ell m'}$ we obtain,
\begin{eqnarray}
H^{(2)}_{mm'} &=& -\frac{1}{2\Delta \epsilon} \sum_\ell  \Delta^2 \left [ \delta_{m,c}\left (\delta_{b\pm1,a} +\delta_{a\pm1,b}  \right ) + \delta_{m,b} \left( \delta_{c\pm 1, a} +\delta_{a\pm1,c}  \right ) \right. \nonumber \\
&&\left.+ \delta_{m,a}\left( \delta_{c\pm1,b} + \delta_{b\pm1,c}\right )\right ] \times \left [  \delta_{c,m'}\left (\delta_{a,b\pm1} +\delta_{b,a\pm1}  \right )  \right. \nonumber \\
&&\left.+ \delta_{b,m'} \left( \delta_{a,c\pm 1} +\delta_{c,a\pm1}  \right )  + \delta_{a,m'}\left( \delta_{b,c\pm1} + \delta_{c,b\pm1}\right )\right ]\nonumber \label{eqn:SW_2_ideal_2} \\ 
\end{eqnarray}
The diagonal elements of $H^{(2)}$ are
\begin{eqnarray}
 H^{(2)}_{mm}  = - \frac{\Delta^2}{\Delta \epsilon} m^* ,
 \label{eqn:SW_2_ideal_diag} 
\end{eqnarray}
where $m^* $ is the total number of nearest-neighbour lattice sites without considering the site $m$. 
The value of $m^*$ can be computed  as,
\begin{eqnarray}
m^* = \sum_{i=0}^{m-1}\sum_{j = i+1}^{m-1} \delta_{i,j\pm1} + \sum_{i=m+1}^{N}\sum_{j = i+1}^{N} \delta_{i,j\pm1},
\end{eqnarray}
where the first summation is over all pairs of lattice site interactions for $i<m$, and the second summation is for $i>m$.
For example, when $m$ is any of a 1D lattice site edges, $m^* =  N-2$, where $N$ is the total number of lattice sites. 
In the case when $m \neq m'$, 
\begin{eqnarray}
H^{(2)}_{mm'} = -\frac{3\Delta^2}{2\Delta \epsilon}    \left [ \delta_{m,m'\pm2}  + \delta_{m\pm2,m'} \right ],\label{eqn:SW_2_ideal_offdiag} 
\end{eqnarray}
which leads to next-nearest-neighbour hopping with $t' = -\frac{3\Delta^2}{2\Delta \epsilon} $.  Combining Eqs. (\ref{eqn:SW_2_ideal_diag})  and (\ref{eqn:SW_2_ideal_offdiag})  we see that the SW transformation, to first order, leads to the following one-particle Hamiltonian:
\begin{eqnarray}
{\cal \tilde{H}} &=&  \sum_{i} \omega'_i \hat{c}^{\dagger}_{i}\hat{c}_{i} +   \sum_{ i } t\left(   \hat{c}^{\dagger}_{i\pm1}\hat{c}_{i}\right ) +   t'\left(  \hat{c}^{\dagger}_{i\pm2}\hat{c}_{i}\right ), 
\end{eqnarray}
where $ \omega'_i =  \Delta \epsilon  - \frac{\Delta^2}{\Delta \epsilon} m^* $ and $t' = -\frac{3  \Delta^2}{2\Delta \epsilon}$.

\end{document}